%% file: main.tex
\documentclass[sigconf]{acmart}
\copyrightyear{2026}
\acmYear{2026}
\setcopyright{cc}
\setcctype{by-nc-nd}
\acmConference[ICPC '26]{34th IEEE/ACM International Conference on Program Comprehension}{April 12--13, 2026}{Rio de Janeiro, Brazil}
\acmBooktitle{34th IEEE/ACM International Conference on Program Comprehension (ICPC '26), April 12--13, 2026, Rio de Janeiro, Brazil}
\acmPrice{}
\acmDOI{10.1145/3794763.3794797}
\acmISBN{979-8-4007-2482-4/2026/04}

\ccsdesc[500]{Software and its engineering~Software testing and debugging}

\input{commands}

\input{macros}
\begin{document}

\title{PALM: Path-aware LLM-based Test Generation with Comprehension}

\author{Yaoxuan Wu}
\orcid{0009-0001-0623-4110}
\affiliation{%
  \institution{UCLA}
  \city{Los Angeles}
  \state{CA}
  \country{USA}
}
\email{thaddywu@cs.ucla.edu}

\author{Xiaojie Zhou}
\orcid{0009-0008-5442-821X}
\affiliation{%
  \institution{UCLA}
  \city{Los Angeles}
  \state{CA}
  \country{USA}
}
\email{xiaojiez6@g.ucla.edu}

\author{Ahmad Humayun}
\orcid{0000-0002-5707-4487}
\affiliation{%
  \institution{Virginia Tech}
  \city{Blacksburg}
  \state{VA}
  \country{USA}
}
\email{ahmad35@vt.edu}

\author{Muhammad Ali Gulzar}
\orcid{0000-0002-8007-8662}
\affiliation{%
  \institution{Virginia Tech}
  \city{Blacksburg}
  \state{VA}
  \country{USA}
}
\email{gulzar@cs.vt.edu}

\author{Miryung Kim}
\orcid{0000-0003-3802-1512}
\affiliation{%
  \institution{UCLA}
  \city{Los Angeles}
  \state{CA}
  \country{USA}
}
\email{miryung@cs.ucla.edu}

\renewcommand{\shortauthors}{Wu et al.}

\input{Text/1.abstract}
\keywords{LLM-assisted testing, symbolic execution}

\maketitle
\input{Text/2.introduction}
\input{Figure/PALM-major-interface}
\input{Text/4.frontend}
\input{Text/5.methodology}
\input{Text/6.evaluations}
\input{Text/7.limitations}
\input{Text/8.relatedwork}
\input{Text/9.conclusion}

\bibliographystyle{ACM-Reference-Format}
\bibliography{reference}
\balance

\end{document}

%% file: commands.tex
\usepackage{multirow}
\usepackage{listings}
\usepackage{pifont}
\usepackage{colortbl}
\usepackage{subcaption}
\usepackage{soul}
\usepackage{makecell}
\usepackage{algorithm}
\usepackage{algorithmic}
\usepackage{xspace}



%% file: macros.tex
\newcommand{\todo}[1]{{\color{purple}{Todo: #1}}}

\newcommand{\code}[1]{{\texttt{#1}}}

\DeclareRobustCommand{\encircle}[1]{%
  \begingroup
  \sbox0{\textcolor{black}{\large \ding{108}}}
  \raisebox{-0.2ex}{%
    \rlap{\usebox0}%
    \hbox to \wd0{%
      \hfil
      \raisebox{0.4ex}{\textcolor{white}{\normalfont\rmfamily\fontsize{7}{7}\selectfont #1}}%
      \hfil
    }%
  }%
  \endgroup
}
\DeclareRobustCommand{\Encircle}[1]{%
  \begingroup
  \sbox0{\textcolor{black}{\large \ding{108}}}
  \raisebox{-0.6ex}{%
    \rlap{\usebox0}%
    \hbox to \wd0{%
      \hfil
      \raisebox{0.4ex}{\textcolor{white}{\normalfont\rmfamily\fontsize{7}{7}\selectfont #1}}%
      \hfil
    }%
  }%
  \endgroup
}

\newcommand{\HLbox}[2]{%
  \begingroup
  \setlength{\fboxsep}{0pt}%
  \colorbox{#1}{\strut\kern0.2pt#2\kern0.2pt}%
  \endgroup
}

\newcommand{\HG}[1]{\HLbox{green!20}{#1}}
\newcommand{\HY}[1]{\HLbox{yellow!20}{#1}}
\newcommand{\btHL}[1]{\HLbox{red!20}{#1}}
\newcommand{\rcomment}[1]{\hfill{\scriptsize // #1}}

%% file: Text/1.abstract.tex
\begin{abstract}
Symbolic execution is a widely used technique for test generation, offering systematic exploration of program paths through constraint solving. However, it is fundamentally constrained by the capability to model the target code including library functions in terms of symbolic constraint and the capability of underlying constraint solvers. As a result, many paths involving complex features remain unanalyzed or insufficiently modeled.
Recent advances in large language models (LLMs) have shown promise in generating diverse and valid test inputs. Yet, LLMs lack mechanisms for systematically enumerating program paths and often fail to cover subtle corner cases. We observe that directly prompting an LLM with the full program leads to missed coverage of interesting paths.

In this paper, we present PALM, a test generation system that combines symbolic path enumeration with LLM-assisted test generation. PALM statically enumerates possible paths through AST-level analysis and transforms each into an executable variant with embedded assertions that specify the target path. This avoids the need to translate path constraints into SMT formulas, by instead constructing program variants that LLM can interpret.
Importantly, PALM provides an interactive frontend that visualizes path coverage alongside generated tests, assembling tests based on the specific paths they exercise. A user study with 12 participants demonstrates that PALM's frontend helps users better understand path coverage and identify which paths are actually exercised by PALM-generated tests, through verification and visualization of their path profiles.

\end{abstract}

%% file: Text/2.introduction.tex
\section{Introduction}
Symbolic execution is a widely adopted software testing and verification technique \cite{king1975new,boyer1975select,king1976symbolic,howden1977symbolic}. It systematically explores different program paths and collects the branch predicates along each path as logical constraints, a.k.a. \textit{path constraints}. These constraints are typically encoded as Satisfiability Modulo Theories (SMT) formulas and passed to constraint solvers for concrete test input generation.

\textit{However, symbolic execution is fundamentally limited by the symbolic encoding required for individual functions and subsequent constraint solving.} As shown in Fig.~\ref{fig:code-example-argparse}, the method \texttt{parseFilePath} processes a list of arguments and attempts to extract the file path following the \texttt{"-f"} flag. A key branch condition in this method involves the string comparison \texttt{args[i].equalsIgnoreCase("-f")}. Constraint solvers such as Z3~\cite{de2008z3} or CVC5~\cite{barbosa2022cvc5} lack built-in support for such operations, unless symbolic execution tool developers manually supply symbolic models for the corresponding library function, such as \texttt{equalsIgnoreCase}.

Recent studies \cite{lemieux2023codamosa, xia2024fuzz4all} have demonstrated that LLMs are very effective in automated test generation. LLMs are capable of understanding program semantics and producing test inputs.

However, LLMs lack the ability to systematically enumerate program paths. As shown in Fig.~\ref{fig:code-example-argparse}, the method \texttt{parseFilePath} parses command-line arguments and extracts the file path following the \texttt{"-f"} flag. GPT-4o generates several test cases that cover common scenarios, such as \texttt{\{"-f","input.txt"\}} and \texttt{\{"-v","-f",\allowbreak "data.csv"\}}. However, it misses the subtle edge case \texttt{\{"-f", "-v"\}}, where the program erroneously interprets \texttt{"-v"} as the file path due to a lack of input validation. The LLM-generated tests fail to expose this bug. We observed that \textit{directly prompting the LLM with the entire program, without indicating which execution paths exist and how to trigger them, often leads to missed edge cases.}


Second, existing LLM-based test generation tools primarily focus on maximizing overall code coverage through automated test generation, but offer limited support for understanding which specific execution paths are exercised by each generated test. To the best of our knowledge, \textit{prior tools lack explicit test–path alignment, limiting the user’s ability to identify missing coverage and guide LLM-driven test generation toward uncovered paths.}

\input{Code/code-example-argparse}
In this paper, we present PALM, a test generation system with comprehension support that combines the path enumeration capability of symbolic execution with the test generation strength of LLMs, while sidestepping the symbolic modeling and constraint solving bottlenecks of traditional symbolic execution. PALM also features an interactive frontend that visualizes the symbolic execution tree, highlights path coverage, and allows users to inspect or refine test generation for specific paths. PALM's test generation consists of two phases:

\paragraph{Path extraction} PALM performs AST-level analysis on the subject program to systematically enumerate program paths. For each path, it automatically constructs a corresponding program variant using program transformation, which is a modified version of the original program. The variant is augmented with \texttt{assertTrue} and \texttt{assertFalse} statements that encode the branch decisions along the path. This variant provides a path-specific prompt that guides the LLM to generate the required test input. In other words, this variant directly serves as the hint for path-specific test generation.

\paragraph{Test generation} 
PALM traverses the symbolic execution tree and generates a test input for each enumerated path. Each generated test is executed on its corresponding variant to verify whether it indeed exercises the target path. The final test suite is organized within the symbolic tree, allowing users to inspect the alignment between test inputs and program paths.

In our experiments, we evaluated PALM on 124 Java programs from HumanEval-Java \cite{jiang2023impact}, which include complex control flows such as nested loops and which use external library functions. By systematically enumerating and extracting program paths, PALM achieves 35.0\% and 24.2\% higher path coverage than GPT-4o-mini and GPT-o3-mini, respectively, when using the same LLM backend. Its iterative test validation and refinement further improves path coverage by 14.2\% over the non-iterative setting with GPT-4o-mini.
In contrast, Symbolic PathFinder (a representative traditional symbolic executor) fails to derive a single accurate path constraint in 34.3\% of the programs due to insufficient symbolic modeling of external API calls.
We also conducted a within-subject user study with 12 participants to understand how much PALM can assist users in generating and comprehending path-aware test inputs. Participants reported higher confidence in their ability to generate sufficient tests, disambiguate redundant or missing tests, and check whether generated tests indeed exercise specific paths, compared to using a pure LLM as is. 

The contributions of PALM are summarized below:
\begin{itemize}
    \item We combine symbolic execution and LLM-assisted test generation by leveraging symbolic execution for systematic path enumeration through AST-level analysis, and utilizing LLMs to generate test inputs without relying on constraint solving.
    \item We design an interactive frontend that visualizes the symbolic execution tree, enabling users to inspect program paths and verify whether LLM-generated tests exercise the intended ones.
    \item We conducted a user study with 12 participants, and the results show that PALM’s interactive frontend helps users better understand path coverage and identify tests that exercise specific paths.
\end{itemize}

%% file: Code/code-example-argparse.tex
\definecolor{codegray}{gray}{0.95}
\definecolor{highlight}{RGB}{255,255,160}
\definecolor{commentgray}{gray}{0.4}

\lstset{
    language=Java,
    basicstyle=\ttfamily\footnotesize,
    keywordstyle=\color{blue},
    stringstyle=\color{orange},
    commentstyle=\color{commentgray}\itshape,
    backgroundcolor=\color{codegray},
    frame=single,
    showstringspaces=false,
    escapeinside=**,
    columns=fullflexible,
    keepspaces=true,
    numbers=left,
    numberstyle=\tiny\color{gray},
    numbersep=8pt,
    breaklines=true,      
    breakatwhitespace=true
}

\begin{figure}[tb]
\scriptsize
\begin{lstlisting}
public class ArgParser {
  public static String parseFilePath(String[] args){
    boolean verbose = false;
    String path = null;
    for (int i = 0; i < args.length; i++) {
        if (args[i].equalsIgnoreCase("-v")) {
            verbose = true;
        }
        else if (args[i].equalsIgnoreCase("-f") 
                    && i + 1 < args.length) {
            path = args[i + 1];
        }
    }
    return path;
  }
}
// GPT-4o generated tests
parseFilePath(new String[]{});
parseFilePath(new String[]{"-v"});
parseFilePath(new String[]{"-f", "input.txt"});
parseFilePath(new String[]{"-v", "-f", "data.csv"});
parseFilePath(new String[]{"-f", "log.txt", "-v"});
parseFilePath(new String[]{"-f"});
\end{lstlisting}
\vspace{-1ex}
\caption{This code snippet parses the argument following \texttt{"-f"} as a file path. While GPT-4o generates tests covering typical cases, it misses the edge case \texttt{\{"-f","-v"\}}, where \texttt{"-v"} is mistakenly interpreted as the file path due to the lack of validation. This path is also challenging for symbolic execution-based testing as SMT solvers like Z3 and CVC5 do not support string operations such as \texttt{equalsIgnoreCase}.}

\Description{A Java code listing for an argument parser that returns the token following \texttt{-f} as the file path, plus a list of GPT-4o-generated test calls. The tests include typical cases such as \{\}, \{\texttt{-v}\}, and \{\texttt{-f}, \texttt{input.txt}\}, but do not include the input \{\texttt{-f}, \texttt{-v}\}.}
\label{fig:code-example-argparse}
\vspace{-3ex}
\end{figure}

%% file: Figure/PALM-major-interface.tex
\begin{figure*}[h]
    \centering
    \includegraphics[width=\linewidth]{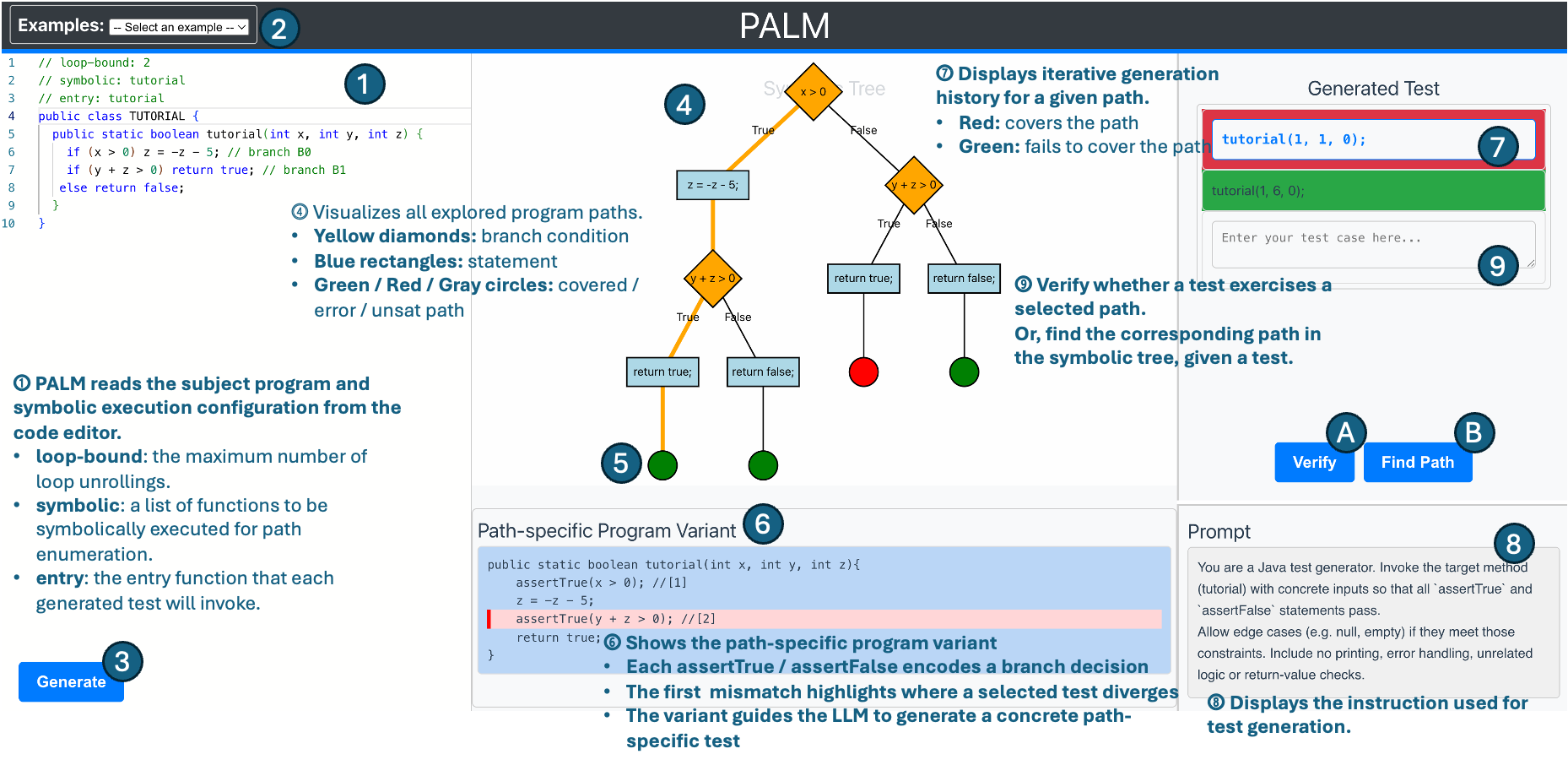}
    \caption{
PALM user interface.
\encircle{1} Code editor and symbolic-execution settings.
\encircle{2} Built-in example selector.
\encircle{3} Start symbolic execution and test generation.
\encircle{4} Symbolic execution tree (leaf nodes show coverage).
\encircle{5} Select a path (click a leaf).
\encircle{6} Path-specific program variant (assertions encode branch decisions).
\encircle{7} Iterative test-generation history.
\encircle{8} Prompt for the selected test.
\encircle{9} Test editor.
\encircle{A} Verify whether a test exercises the selected path.
\encircle{B} Locate the corresponding path for a given test.
}

\Description{A screenshot of the PALM user interface annotated with numbered callouts. The interface includes a code editor and symbolic-execution settings, an example selector, a button to start symbolic execution and test generation, a symbolic execution tree with leaf nodes indicating coverage, a panel to select a path and view the corresponding path-specific program variant with embedded assertions, a history of iterative test-generation attempts, a prompt panel for the selected test, and a test editor. Additional controls allow verifying whether a test exercises the selected path and locating the path corresponding to a given test.}
    \label{fig:PALM-major-interface}
\end{figure*}

%% file: Text/4.frontend.tex
\section{Tool Overview}

PALM is a test generation system for Java programs that combines symbolic execution with LLMs. For each enumerated path, PALM uses an LLM to generate a path-exercising test input. Its interface centers around a symbolic execution tree, a tree-based representation where each node corresponds to a program statement or a branch condition, and each path from the root to a leaf represents a possible execution path. The symbolic tree visualizes explored paths and highlights covered ones in green. Users can visualize the execution trace of a given test and align test inputs with path-level program execution behavior. In the following three subsections, we explain PALM's key features using the example shown in Fig.~\ref{fig:PALM-major-interface}.

\subsection{Test generation}

At its core, PALM systematically enumerates program paths and invokes an LLM to generate test inputs for each enumerated path. Users begin by providing the subject program, such as a program named \textit{tutorial} in the example figure (\encircle{1}), along with symbolic execution configurations. In this example, the method \textit{tutorial} is marked as the entry point and symbolic, meaning its internal paths will be enumerated. Although the program contains no loops, users can still configure a loop bound, such as 2, to control the loop unrolling depth in programs. After clicking the Test Generation Button (\encircle{3}), PALM begins enumerating paths, constructing program variants, and requesting LLMs for path-specific test generation.

\subsection{Path understanding with symbolic tree and path variant}

An interactive symbolic execution tree serves as a central component in PALM. This tree visualizes the explored execution paths of the program and distinguishes branch conditions (yellow diamond nodes) from regular statements (blue rectangular nodes). It also encodes path coverage status as green (covered), red (uncovered), and gray (unreachable). For example, the highlighted orange path (\encircle{5}) contains two conditional branches, \texttt{x>0} and \texttt{y+z>0}, both of which must evaluate to \texttt{true}. The non-conditional assignment \texttt{z=-z-5} appears between them and affects the outcome of the second condition, illustrating a data dependency along the path.

Once a path is selected, PALM presents its corresponding \textit{path-specific program variant} as an executable program (\encircle{6}). This variant is constructed from the symbolic execution trace, augmented with \texttt{assertTrue} or \texttt{assertFalse} statements that reflect the outcome of each branch condition along the path. In the above example, since both conditions must evaluate to true, the variant contains \texttt{assertTrue(x>0)} and \texttt{assertTrue(y+z>0)}.

The variant not only helps users interpret the logic of an execution path, but also serves as a prompt that guides the LLM to generate test inputs satisfying the conditions along that path. Unlike traditional symbolic execution tools that translate path constraints into SMT formulae, PALM leverages LLMs to handle complex language features, such as string operations and library calls, that are often inexpressible in SMT logic.
This approach eliminates the need for manually modeling library functions or treating them as uninterpreted, a key advantage over conventional symbolic executors.

\vspace{-2ex}
\subsection{Validating and refining generated tests}

LLMs may generate a concrete test input that diverges from the intended path. After receiving the LLM generated test, PALM verifies whether it exercises the selected path. When a path is selected (\encircle{5}), PALM displays the generation history for that path. Each test is color-coded: green boxes indicate successful coverage (e.g., \texttt{tutorial(1,6,0)} covers the highlighted path), while red boxes denote failures (e.g., \encircle{7} \texttt{tutorial(1,1,0)} diverges at the second branch \texttt{y+z>0}). When a test diverges from the selected path, PALM highlights the first failing assertion in the path variant (\encircle{6}); in this example, it is \texttt{assertTrue(y+z>0)}.

Users can inspect the divergence and hypothesize the cause. For example, the LLM might ignore the update \texttt{z=-z-5} before evaluating the \texttt{y+z>0} branch. To guide the LLM, users can manually edit the prompt in the Prompt Box (\encircle{8}) and inject guidance such as: \textit{"The y+z$>$0 branch should consider that z is updated as -z-5."} Users can also update the test to \texttt{tutorial(1,6,0)} and verify whether it covers the intended path (\encircle{A}), or identify the path actually taken by a test (\encircle{B}).

In PALM's feedback-based refinement loop, the system provides the previous failing test and the first diverging assertion to the LLM to guide subsequent attempts. This helps the model pinpoint where its prior generation deviated from the target path. By default, PALM performs up to five iterations per path to produce a valid concrete path-covering test.


%% file: Text/5.methodology.tex
\input{Figure/workflow}

\section{Methodology}

PALM's backend test generation consists of two phases: (1) path enumeration and extraction and (2) path-specific concrete test generation with validation. Fig.~\ref{fig:workflow} presents an overview of this workflow.

PALM takes a Java program as input and systematically enumerates its execution paths using AST-level analysis. For each enumerated path, PALM generates a corresponding path-specific program variant, which is an executable program that represents the statements and branch decisions along the path. These variants encode each path's semantics in a form of normal Java statements that is interpretable by LLMs, without translating to SMT constraints.
To facilitate precise control over test generation, PALM organizes and visualizes paths into a symbolic execution tree.

In the second phase, PALM uses the constructed path-specific program variants as prompts to guide LLMs in generating concrete test inputs. Each generated test is dynamically validated against its corresponding program variant to ensure that it indeed exercises the intended execution path. When validation fails, PALM provides the LLM with feedback about the failed assertion to iteratively refine the test.

\subsection{Path Extraction}

PALM conducts path enumeration at the AST level and transforms each path into a program variant. Each variant encodes the evaluation outcomes of all selected branches using \code{assertTrue} and \code{assertFalse} statements, thereby preserving the semantics of the execution path.

\input{Figure/code-example-ispalindrome}

Fig.~\ref{fig:code-example-ispalindrome} illustrates an example of path extraction. The subject program performs a palindrome check using a loop with an inner conditional. The extracted path (c) corresponds to an execution trace where the loop condition \texttt{i<len} must hold initially (\encircle{2}), and the inner branch condition \texttt{text.charAt(i) != text.charAt(len-i-1)} must evaluate to \texttt{true} (\encircle{3}). In contrast, path (d) corresponds to a different trace where the loop conditions must be true in two consecutive iterations (\encircle{2}, \encircle{5}); the inner branch must evaluate to \texttt{false} in the first iteration (\encircle{4}), and \texttt{true} in the second iteration (\encircle{6}).

Algorithm~\ref{alg:psuedo-path-extract} formalizes our path-enumeration process. It recursively traverses the program’s AST and compositionally constructs the set of enumerated paths. Lines 2–5 handle \code{If-Then-Else} constructs by splitting into the two branches and inserting \code{assertTrue} or \code{assertFalse} based on the branch outcome. For example, \encircle{4} and \encircle{6} correspond to the two branches of the inner conditional check \texttt{text.charAt(i) != text.charAt(len-i-1)} in Fig.~\ref{fig:code-example-ispalindrome}.

Lines 6–8 handle \code{While} loops by unrolling the loop body up to a user-defined loop-unrolling bound $K$. Other loop constructs such as \code{Do-While} and \code{For} loops are handled analogously. Line 10 processes compound \code{BlockStatement} structures consisting of sequential statements by enumerating the combinations of subpaths induced by each statement.

\input{Code/extraction}

A common practice in symbolic execution is to allow annotations specifying which functions should undergo path enumeration (referred to as symbolic functions), as well as which function serves as the entry point. This allows users to focus on path exploration on functions of interest. PALM supports this flexibility as well.

Since path extraction flattens control structures and isolates a single execution path per symbolic function, key semantic information may be lost if not preserved explicitly. To address this, we apply function inlining and variable renaming to preserve the semantics of nested calls and variable scopes. Additionally, we perform constant propagation and folding to simplify each path-specific program variant, reducing syntactic complexity and easing LLM reasoning.

\input{Code/function-inlining-algorithm}

\paragraph{Function inlining}
A symbolic function may invoke other symbolic functions multiple times, each leading to a different execution path. If we retain only a single path-specific variant per function, this can lead to incomplete path semantics. For example, suppose symbolic function \texttt{A} calls another symbolic function \texttt{B} twice, and \texttt{B} takes different paths in each call. Without inlining, we would need to represent \texttt{B} with multiple variants, complicating test generation.

To address this, PALM inlines all symbolic callees within each caller’s path variant. This ensures that the semantics of nested function calls are fully embedded in the caller’s path, allowing PALM to construct a flattened, self-contained path variant for the entry function. Consequently, the final prompt to the LLM contains a single expanded version of the entry function with all relevant behaviors explicitly included.

Algorithm~\ref{alg:function-inlining-algorithm} formalizes this process. PALM first extracts intra-procedural path variants for all annotated symbolic functions. For each callsite in the entry function, PALM enumerates all corresponding callee paths and replaces the call with each of those inlined bodies. This ensures that all possible call behaviors are embedded and the full execution semantics are preserved, without requiring multiple variant copies of the same symbolic function.

\paragraph{Variable renaming}
In the original subject program, variable scopes are managed implicitly by the nesting structure. However, after path extraction with control flow expansion, such as loop unrolling or function inlining, variable scopes are flattened. As a result, variable reuse can lead to name conflicts or ambiguity. To preserve semantic clarity and avoid confusion during test generation, PALM renames local variables by appending a numeric suffix.

For example, if a variable \texttt{i} is declared inside a loop and the loop is unrolled twice, the two versions will be renamed as \texttt{i\_0} and \texttt{i\_1}, explicitly reflecting their different iterations.

\paragraph{Constant propagation and folding}
PALM applies intra-procedural constant propagation and constant folding. This step is performed after function inlining and variable renaming, and it only propagates numerical and boolean constants. For example, in Fig.~\ref{fig:code-example-ispalindrome}, \encircle{2} contains \texttt{assertTrue(i<len);} where \texttt{i} is initialized to \texttt{0}; this can be folded into \texttt{assertTrue(0<len)}. Constant folding also enables pruning of trivially infeasible paths, such as \texttt{assertTrue(x < y)} where \texttt{x=1} and \texttt{y=0}. These simplifications reduce the syntactic complexity of path variants, easing both the LLM’s symbolic reasoning and human understanding of path semantics.

\subsection{Path Aware Test Generation with Validation}

Inspired by traditional symbolic execution, PALM explores paths in a depth-first order, conducting feasibility checks during exploration.

\paragraph{Test generation with a symbolic execution tree} After extracting path variants, PALM organizes them into a symbolic execution tree, where each leaf node corresponds to a complete path variant, and internal nodes represent shared prefixes between paths. PALM performs test generation by traversing this tree in a top-down fashion. Algorithm~\ref{alg:test-generation} formalizes this process. Line 7-12 send the corresponding path variant to the LLM for test generation. For each path, PALM allows up to five trials to generate an assertion-satisfying test and backtracks if the path is deemed infeasible. This symbolic execution tree not only facilitates test generation, but also provides the foundation for checking whether a specific path is covered and for visualizing test coverage at the path level.

\input{Code/symbolic-tree-algorithm}

\input{Table/PALM-prompt}

When constructing the prompt for test generation, PALM retains the full implementation of class fields and non-symbolic methods from the input program. This provides essential context for the LLM to reason about (1) input-output behavior and (2) internal state transitions within non-symbolic functions. PALM excludes the implementation of external or third-party library functions not present in the input code, as modern LLMs are already well-trained on common APIs. Including such definitions would be redundant and could dilute prompt clarity by increasing its length without contributing additional semantic value. Table \ref{tab:PALM-prompt} shows PALM's prompt template during test generation.

By leveraging LLMs for test generation, PALM eliminates the need to symbolically model used library functions, a major pain point of traditional symbolic execution engines such as Symbolic Pathfinder~\cite{puasuareanu2010symbolic}. In such engines, developers must either manually encode each function’s semantics as symbolic constraints or treat them as uninterpreted.

\paragraph{Path exercisability validation via runtime execution}
When the LLM produces a test, PALM runs it on the corresponding program variant to check whether it exercises the target path. If it fails, PALM reports the first violated assertion to the LLM for refinement.
For example, for the path in Fig.~\ref{fig:code-example-ispalindrome}(d), the LLM may generate \texttt{is\_palindrome("ab")}. The run violates the assertion at \encircle{6} (the second iteration should evaluate the condition to \texttt{true}). PALM feeds back \texttt{assertTrue(text.charAt(i) != text.charAt(len-i-1))}, and the LLM corrects the test to \texttt{is\_palindrome("abca")}, which exercises the intended path.

%% file: Figure/workflow.tex
\begin{figure*}[tb]
    \centering
    \includegraphics[width=0.92\linewidth]{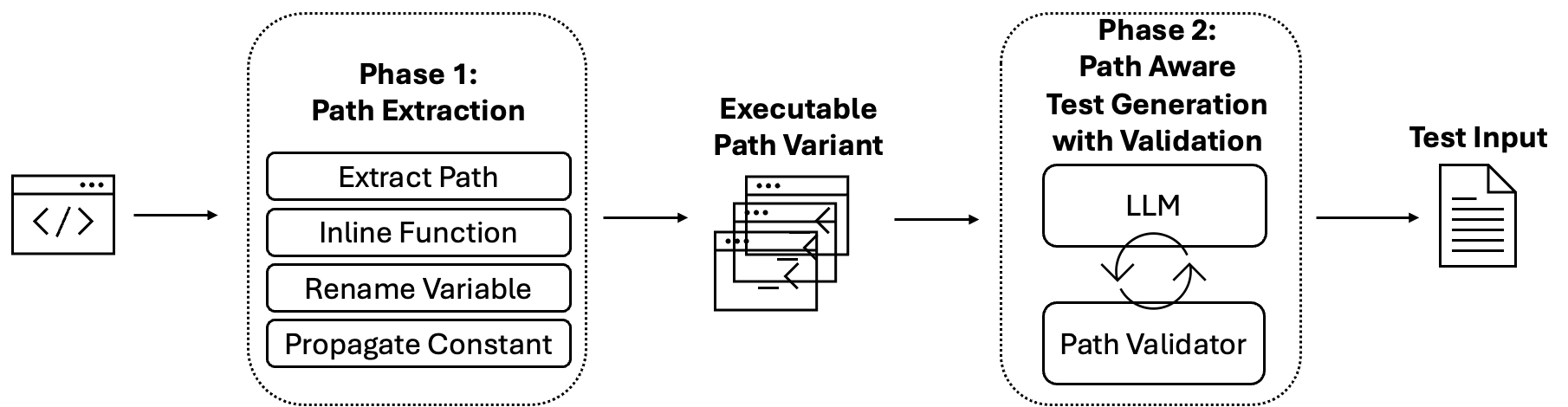}
    \caption{PALM has two phases: (1) enumerate paths and synthesize executable path-specific variants via loop unrolling, inlining, renaming, and constant propagation/folding; (2) traverse the path tree, call an LLM to generate inputs, and iteratively validate them by execution, regenerating with feedback when a test misses the intended path.}
    
    \Description{A workflow diagram of PALM with two phases. The first phase enumerates program paths and constructs executable path-specific variants using transformations such as loop unrolling, inlining, renaming, and constant propagation and folding. The second phase traverses the resulting path tree, queries an LLM to generate test inputs for selected paths, executes the variants to validate whether the tests follow the intended path, and iteratively regenerates tests using feedback when validation fails.}
    \label{fig:workflow}
\end{figure*}

%% file: Figure/code-example-ispalindrome.tex
\definecolor{codegray}{gray}{0.95}
\definecolor{highlight}{RGB}{255,255,160}
\definecolor{commentgray}{gray}{0.4}

\lstset{
    language=Java,
    basicstyle=\ttfamily\footnotesize,
    keywordstyle=\color{blue},
    stringstyle=\color{orange},
    commentstyle=\color{commentgray}\itshape,
    backgroundcolor=\color{codegray},
    frame=single,
    showstringspaces=false,
    escapeinside=**,
    columns=fullflexible,
    keepspaces=true,
    numbers=left,
    numberstyle=\tiny\color{gray},
    numbersep=8pt,
    breaklines=true,      
    breakatwhitespace=true
}

\begin{figure*}[t]
\centering

\begin{minipage}[t]{0.48\textwidth}
\lstset{numbers=left}
\begin{lstlisting}
// symbolic: is_palindrome
// entry: is_palindrome
public static boolean is_palindrome(String text) {
  int len = text.length();
  for (int i = 0; *\HG{i < len}*; i += 1) {
    if (*\HY{text.charAt(i) != text.charAt(len-i-1)}*)
      return false;
  }
  return true;
}
\end{lstlisting}
\vspace{-2.6ex}
\caption*{(a) Program under test: palindrome test}
\end{minipage}
\hfill
\begin{minipage}[t]{0.48\textwidth}
\lstset{numbers=left}
\begin{lstlisting}
public static boolean is_palindrome(String text) {
  int len = text.length();
  int i = 0;
  *\HG{assertTrue(i < len);}* *\Encircle{2}* // [1]
  *\HY{assertTrue(text.charAt(i) !=}* // [2]
              *\HY{text.charAt(len-i-1));}* *\Encircle{3}*
  return false;
}
\end{lstlisting}
\vspace{-2.5ex}
\caption*{(c) Path 2's program variant}
\end{minipage}

\vspace{1em}

\begin{minipage}[t]{0.48\textwidth}
\vspace*{\fill}
\centering
\includegraphics[width=0.7\textwidth]{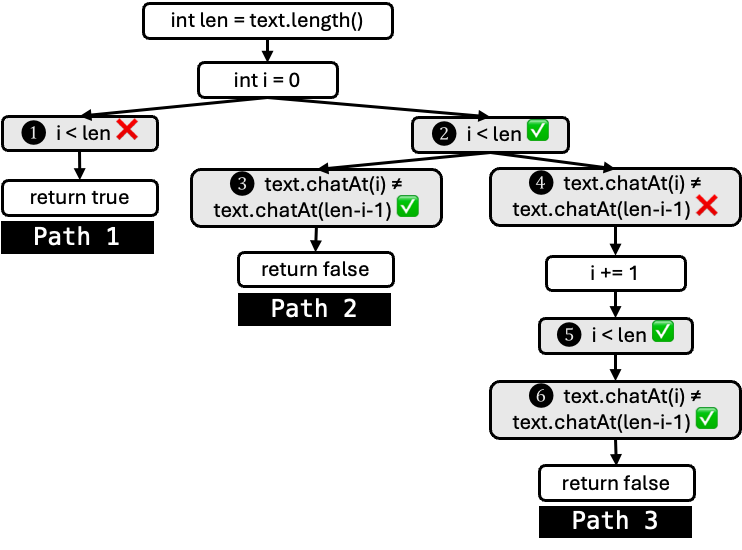}
\vspace{-2.5ex}
\caption*{(b) Symbolic tree}
\end{minipage}
\hfill
\begin{minipage}[t]{0.48\textwidth}
\lstset{numbers=left}
\begin{lstlisting}
public static boolean is_palindrome(String text) {
  int len = text.length();
  int i = 0;
  *\HG{assertTrue(i < len);}* *\Encircle{2}* // [1]
  *\HY{assertFalse(text.charAt(i) !=}* // [2]
              *\HY{text.charAt(len-i-1));}* *\Encircle{4}*
  i += 1;
  *\HG{assertTrue(i < len);}* *\Encircle{5}* // [3]
  *\HY{assertTrue(text.charAt(i) !=}* // [4]
              *\HY{text.charAt(len-i-1));}* *\Encircle{6}*
  return false;
}
\end{lstlisting}
\vspace{-2.5ex}
\caption*{(d) Path 3's program variant}
\end{minipage}

\caption{Example palindrome checker with a \texttt{for} loop and two branches (green, yellow). (b) shows the symbolic execution tree, where each edge corresponds to a source predicate. After loop unrolling, PALM derives concrete path conditions as assertion-based encodings in (c) and (d), translating each branch to an \texttt{assertTrue}/\texttt{assertFalse} over the predicate. For example, \Encircle{3} indicates the yellow predicate evaluates to \texttt{true} in the first iteration.}

\Description{A four-part figure illustrating PALM on a palindrome checker. (a) shows a Java method \texttt{is\_palindrome} with a \texttt{for} loop and a character-comparison branch. (b) shows the corresponding symbolic execution tree with edges labeled by the loop condition and the branch predicate. (c) and (d) show two path-specific program variants produced after loop unrolling, where the loop and branch predicates are encoded as \texttt{assertTrue} and \texttt{assertFalse} statements to represent different execution paths across iterations.}
\label{fig:code-example-ispalindrome}
\end{figure*}

%% file: Code/extraction.tex
\begin{algorithm}
\caption{Path Extraction for each Symbolic Function}
\label{alg:psuedo-path-extract}
{\scriptsize
\begin{algorithmic}[1]
\REQUIRE Program $program$, loop bound $K$ (default = 2)
\ENSURE $Paths_m$: a set of paths for symbolic function $m$
\STATE \textbf{Procedure} \texttt{ExtractPath}($node$) \rcomment{$node$ is an AST node}
\IF{$node$ is \texttt{IfStatement}}
  \STATE $P_{\text{then}} \gets \texttt{PrependAll}(\texttt{assertTrue}(node.branch), \texttt{ExtractPath}(node.thenBlock))$
  \STATE $P_{\text{else}} \gets \texttt{PrependAll}(\texttt{assertFalse}(node.branch), \texttt{ExtractPath}(node.elseBlock))$
  \STATE \textbf{return} $P_{\text{then}} \cup P_{\text{else}}$
\ELSIF{$node$ is \texttt{WhileStatement}}
  \STATE Unroll the loop for 0 to $K$ iterations, then apply \texttt{ExtractPath} to each unrolled version
  \STATE \textbf{return} the union of all extracted paths
\ELSIF{$node$ is \texttt{BlockStatement}}
  \STATE \textbf{return} combinations of \texttt{ExtractPath}($s$), for all $s \in node$
\ENDIF
\STATE $\dots$ \rcomment{Handle other code structures}
\STATE \textbf{End Procedure}
\end{algorithmic}
}
\end{algorithm}

%% file: Code/function-inlining-algorithm.tex
\begin{algorithm}[h]
\caption{Function Inlining}
\label{alg:function-inlining-algorithm}
{\scriptsize
\begin{algorithmic}[1]
\REQUIRE $P_m$: set of symbolic paths for each symbolic function $m$
\REQUIRE $R$: recursion bound (default = 2)
\ENSURE $T_m$: complete set of inlined paths for function $m$

\STATE $T \gets$ paths without symbolic calls
\STATE $N \gets$ paths with symbolic calls
\FOR{recursion level $r = 1$ to $R$}
  \STATE $N' \gets \emptyset$
  \FOR{each non-terminating path $n \in N$}
    \STATE Identify call-sites $c_1, \dots, c_k$ in $n$ calling functions $m_1, \dots, m_k$
    \STATE Enumerate all path tuples $(p_1, \dots, p_k)$ where $p_i \in T_{m_i}$
    \STATE $N' \gets N' \cup$ the set of paths obtained by replacing each $c_i$ in $n$ with $p_i$
  \ENDFOR
  \STATE $T \gets T \cup N'$ \rcomment{Prepare for next inlining level}
  \STATE $N \gets N'$ \rcomment{Continue inlining remaining calls}
\ENDFOR
\STATE \textbf{return} $T$
\end{algorithmic}
}
\end{algorithm}

%% file: Code/symbolic-tree-algorithm.tex
\begin{algorithm}[b]
\caption{Test Generation on Symbolic Tree}
\label{alg:test-generation}
{\scriptsize
\begin{algorithmic}[1]
\REQUIRE $root$: the root node of the program's symbolic execution tree
\ENSURE $T$: a test suite aligned with program paths

\STATE \textbf{Procedure} \texttt{GenerateTests}($node$)
\STATE $p \gets$ the path condition from $root$ to $node$
\IF{$node$ is a leaf}
  \STATE $t \gets$ attempt to generate a test for path $p$ (up to 5 trials)
  \IF{$t$ is not found}
    \STATE \textbf{return} $\emptyset$
  \ENDIF
  \STATE \textbf{return} $\{t\}$ \rcomment{Return a valid test if exists}
\ENDIF
\IF{$node$ is divergent}
  \STATE Run test generation for path $p$
  \IF{unsatisfiable or generation fails after 5 trials}
    \STATE \textbf{return} $\emptyset$
  \ENDIF
\ENDIF
\STATE \textbf{return} $\bigcup \texttt{GenerateTests}(c)$, $\forall c \in child(node)$
\STATE \textbf{End Procedure}
\end{algorithmic}
}
\end{algorithm}

%% file: Table/PALM-prompt.tex
\begin{table}[t]
\centering
\caption{PALM prompt template. Path constraints are encoded as \texttt{assertTrue}/\texttt{assertFalse} statements, and failures provide assertion-level feedback for iterative refinement.}
\Description{A table showing the PALM prompt template with four labeled sections: (1) Task Description instructing the LLM to generate Java tests that satisfy all \texttt{assertTrue} and \texttt{assertFalse} statements without printing or extra logic; (2) Input program listing placeholders for \texttt{[imports]}, \texttt{[other functions]}, and the \texttt{[path condition]}; (3) Generation History recording the current round number, previous \texttt{[test code]} and any \texttt{[failed assertion]} used for iterative refinement; and (4) Output Format specifying the expected test output. Placeholders shown in the table (for example \texttt{[focal method]}) are meant to be substituted when constructing a path-specific prompt.}

\label{tab:PALM-prompt}
\vspace{-2ex}
\begin{tabular}{p{\linewidth}}
\toprule

\rowcolor{gray!30}
Task Description \\
You are a Java test generator. Invoke the target method \textbf{\colorbox{yellow!30}{[focal method]}} with concrete inputs so that all \texttt{assertTrue} and \texttt{assertFalse} statements pass. Allow edge cases (e.g., \texttt{null}, empty) if they meet those constraints. Include no printing, error handling, unrelated logic, or return-value checks. \\

\rowcolor{gray!30}
Input program \\
\textbf{\colorbox{yellow!30}{[imports]}}
\textbf{\colorbox{yellow!30}{[other functions]}}
\textbf{\colorbox{yellow!30}{[path condition]}} \\

\rowcolor{gray!30}
Generation History \\
Round \textbf{\colorbox{yellow!30}{[i]}} generation: \textbf{\colorbox{yellow!30}{[test code]}}. \\
Failed assertion: \textbf{\colorbox{yellow!30}{[failed assertion]}}. \\

\rowcolor{gray!30}
Output Format \\
\textbf{\colorbox{yellow!30}{[output instruction]}} \\

\bottomrule
\end{tabular}
\end{table}

%% file: Text/6.evaluations.tex
\section{Evaluation}

We answered the following research questions:
\begin{itemize}
    \item \textbf{RQ1:} Does PALM's path-aware program variant generation improve path coverage?
    \item \textbf{RQ2:} Does PALM's path exercisability validation and error-feedback guided test generation improve path coverage?
    \item \textbf{RQ3:} Does PALM overcome the limitation of traditional symbolic executors (Symbolic Pathfinder) on handling external functions without symbolic modeling?
    \item \textbf{RQ4:} Does PALM help better understand and identify redundant or missing test cases in terms of different path profiles and produce concrete tests aligned with specific paths?
\end{itemize}

\paragraph{Dataset} We used 124 benchmark programs from HumanEval-Java, a Java benchmark for evaluating automated program repair tools \cite{jiang2023impact}. Its programs contain various external method calls, data structure manipulations, and complex control flow.

\paragraph{Implementation} We implemented the path extraction component based on \textsc{JavaParser} \cite{javaparser}, and developed the frontend using \textsc{Vue}.

\paragraph{Configuration} In our evaluation, we set the loop-unrolling and recursion-depth bounds to 2. We treat the method whose name matches the target class as the symbolic entry, and treat all other methods (including program-defined helpers and external API calls) as non-symbolic and excluded from path enumeration. During test generation, we enumerate up to 50 distinct paths per program (including infeasible paths), capping exploration to mitigate path explosion and bound analysis cost. We chose 50 as a practical budget that keeps the total cost bounded while covering the majority of feasible paths in HumanEval-Java. Increasing this limit leads to diminishing returns in path coverage relative to the additional LLM cost. Each experiment is run 5 times.

\input{Text/6.evaluations-LLM}
\input{Text/6.evaluations-refine}
\input{Text/6.evaluations-SPF}
\input{Text/6.evaluations-userstudy}

%% file: Text/6.evaluations-LLM.tex
\subsection{Path-aware test generation}

\input{Table/evaluation-LLM}

In this section, we evaluate the effectiveness of PALM's path-aware test generation strategy and compare to LLM's capability to generate tests. Fig. \ref{fig:code-example-anyint} shows the code snippet of \texttt{any\_int}. LLM-generated tests fail to cover the highlighted else-branch, corresponding to the edge case where some inputs are non-integers. In contrast, PALM systematically enumerates program paths and generates an input \texttt{any\_int(3.0,1.1,2.0)} to cover this branch.

\input{Code/code-example-anyint}

Table \ref{tab:evaluation-LLM} illustrates PALM's code coverage against LLM. PALM with GPT-4o-mini achieves a 35.0\% higher path coverage, 5.9\% branch coverage, 4.0 \% line coverage than direct LLM generation with GPT-4o-mini. It is worth noting that PALM with GPT-4o-mini achieves higher path coverage than direct LLM generation with a stronger model GPT-o3-mini. PALM with GPT-o3-mini achieves a 24.2\% higher path coverage, 1.7\% lower branch coverage, 0.4\% lower line coverage than direct LLM generation with GPT-o3-mini.

We also observe that, as the LLM backend becomes stronger, direct generation improves and can even outperform PALM under our current configuration: with GPT-o4-mini, direct generation achieves 8.8\%, 1.1\%, and 0.4\% higher path, branch, and line coverage than PALM, respectively. This is partly because PALM caps exploration at 50 enumerated paths per program to mitigate path explosion. Nevertheless, PALM remains complementary via path-level guidance and test-path visualization for diagnosing missing or redundant coverage.



%% file: Table/evaluation-LLM.tex
\begin{table}[t]
\centering
\caption{Comparison of average test coverage between PALM and LLM under different LLM backends. Each configuration was executed five times to reduce variance. Reported cost represents the total for all 124 benchmarks.}
\Description{A table comparing average test coverage and total cost for PALM and a direct LLM baseline across three LLM backends: GPT-4o-mini, GPT-o3-mini, and GPT-o4-mini. For each backend, the table reports average path coverage, branch coverage, and line coverage, along with the total dollar cost for running all 124 benchmarks, with each configuration repeated five times.}
\label{tab:evaluation-LLM}
\renewcommand{\arraystretch}{1.2}
\begin{tabular}{>{\bfseries}p{1.5cm} p{1.5cm} p{1.5cm} p{1.3cm} p{1cm}}
\toprule
\rowcolor{gray!30}
Tool & Path Cov & Branch Cov & Line Cov & Cost (\$) \\
\midrule
\rowcolor{gray!20}
\multicolumn{5}{c}{\textbf{w/ GPT-4o-mini}} \\
PALM & 663.0 & 198.5 & 948.5 & 0.58 \\
LLM  & 491.5 & 187.4 & 912.1 & 0.15 \\
\midrule
\rowcolor{gray!20}
\multicolumn{5}{c}{\textbf{w/ GPT-o3-mini}} \\
PALM & 775.8 & 199.4 & 950.6 & 8.52 \\
LLM  & 624.6 & 202.8 & 954.8 & 0.63 \\
\midrule
\rowcolor{gray!20}
\multicolumn{5}{c}{\textbf{w/ GPT-o4-mini}} \\
PALM & 746.2 & 201.0 & 955.2 & 6.14 \\
LLM  & 811.6 & 203.2 & 959.0 & 0.74 \\
\bottomrule
\end{tabular}
\end{table}

%% file: Code/code-example-anyint.tex
\definecolor{codegray}{gray}{0.95}
\definecolor{highlight}{RGB}{255,255,160}
\definecolor{commentgray}{gray}{0.4}

\lstset{
    language=Java,
    basicstyle=\ttfamily\footnotesize,
    keywordstyle=\color{blue},
    stringstyle=\color{orange},
    commentstyle=\color{commentgray}\itshape,
    frame=none,                 
    backgroundcolor={},         
    showstringspaces=false,
    escapeinside=**,
    columns=fullflexible,
    keepspaces=true,
    numbers=left,
    numberstyle=\tiny\color{gray},
    numbersep=8pt,
    breaklines=true,
    breakatwhitespace=true
}

\newsavebox{\codeboxanyint}

\begin{figure}
\centering

\setlength{\fboxsep}{3pt}     
\setlength{\fboxrule}{0.4pt}  

\begin{lrbox}{\codeboxanyint}
\begin{minipage}{0.98\linewidth}
\begin{lstlisting}
public static boolean any_int(double x, double y, double z) {
  if ((int) x == x && (int) y == y && (int) z == z) {
    if (x + y == z || x + z == y || y + z == x)
      return true;
  } *\btHL{// else-branch}*
  return false;
}
\end{lstlisting}
\end{minipage}
\end{lrbox}

\fcolorbox{black}{codegray}{\usebox{\codeboxanyint}}

\vspace{-1ex}
\caption{Code snippet from \texttt{any\_int}. The condition \texttt{(int)x == x} checks whether \texttt{x} is an integer (i.e., has no fractional part). The highlighted else-branch corresponds to inputs where at least one of \texttt{x}, \texttt{y}, or \texttt{z} is not an integer. LLM-generated tests (GPT-4o-mini) fail to cover this branch, whereas PALM covers with \texttt{any\_int(3.0,1.1,2.0)}.}
\Description{A formatted Java code listing for the method \texttt{any\_int(double x, double y, double z)}. The code first checks whether \texttt{x}, \texttt{y}, and \texttt{z} have no fractional part using comparisons such as \texttt{(int)x == x}, and then tests whether any pairwise sum equals the third value. An \texttt{else}-branch is highlighted to indicate the case where at least one input is not an integer, after which the method returns \texttt{false}.}
\label{fig:code-example-anyint}
\vspace{-2ex}
\end{figure}

%% file: Text/6.evaluations-refine.tex
\subsection{Test validation and refinement}

\input{Figure/evaluation-progress-tex}

Fig.~\ref{fig:evaluation-progress} shows the path, branch, and line coverage achieved by iteratively generated tests over $K$ rounds. With the weaker model GPT-4o-mini, four additional rounds of test generation lead to substantial improvements: 14.2\% in path coverage, 4.2\% in branch coverage, and 1.4\% in line coverage. The coverage gains plateau after the fifth round, indicating convergence. In contrast, the stronger reasoning model GPT-o3-mini typically achieves high coverage for each path on the first attempt, with only marginal improvements across rounds: 0.5\% in path coverage, 0.5\% in branch coverage, and 0.1\% in line coverage.

%% file: Figure/evaluation-progress-tex.tex
\begin{figure}[t]
\centering
\includegraphics[width=\linewidth]{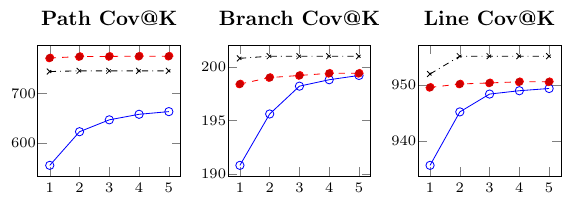}

\caption{Test coverage progress of PALM with $k$ rounds of trial for each program path using three LLM backends}
\Description{A line chart showing test coverage progress of PALM as the number of trial rounds $k$ increases for each program path. The chart compares three LLM backends, with separate curves illustrating how coverage improves over successive rounds for each backend.}
\label{fig:evaluation-progress}
\vspace{-1ex}
\end{figure}

%% file: Text/6.evaluations-SPF.tex
\subsection{Comparison with Symbolic PathFinder}

\input{Table/SPF-analysis}
We compared PALM against Symbolic PathFinder (SPF). As is typical for many symbolic execution engines, SPF requires symbolic modeling of used API functions or considers them as uninterpreted.

\input{Code/code-example-split}
Fig.~\ref{fig:code-example-split} shows an example, where SPF fails to model the path constraints due to two unsupported API calls and to model a symbolic heap object.
Symbolic modeling is particularly challenging here because SPF would need to represent the heap object returned by \texttt{split()}, namely a \texttt{String[]} array, a type that SPF does not currently support symbolically. 
Alternatively, falling back to concrete execution on \texttt{split()} loses the dependency between \texttt{txt} and \texttt{parts}, resulting in an incomplete path modeling.

Table~\ref{tab:SPF-analysis} summarizes the list of library functions that have not been modeled by SPF. As a result, SPF fails to model even a single complete path in 49 programs (34.3\%). In contrast, PALM does not lower path constraints into SMT.
PALM adopts symbolic execution-style path enumeration and utilizes LLM to infer constraints such as \texttt{parts=txt.split(" ")}, bypassing the need for explicit symbolic modeling.

%% file: Table/SPF-analysis.tex
\newcolumntype{T}[1]{>{\ttfamily\arraybackslash}p{#1}}

\begin{table*}[t]
\centering
\caption{Java code patterns that lack symbolic support in Symbolic PathFinder (SPF). For these cases, SPF either terminates prematurely or falls back to concrete execution without complete path modeling.}
\Description{A table summarizing Java code patterns that lack symbolic support in Symbolic PathFinder (SPF). The rows list four categories—String operations, type conversion, generic types, and other library calls—each with example code structures, the number and percentage of affected benchmark programs, and a brief explanation of why SPF fails (for example, missing symbolic models for String APIs such as \texttt{split()}, lack of SMT support for string-to-double conversion, difficulty modeling generic-object methods such as \texttt{Object.equals} and \texttt{Object.compareTo}, and fallback to concrete execution for unsupported library calls).}
\label{tab:SPF-analysis}

\renewcommand{\arraystretch}{1.2}
\begin{tabular}{>{\bfseries}p{3cm} T{3.5cm} p{2.8cm} p{6.7cm}}
\toprule
\rowcolor{gray!20}
Category & Example Code Structure & \# Affected Programs & SPF Failure Reason \\
\midrule
String Operations &
split(), toCharArray(), toLowerCase() &
39 (27.3\%) &
Lack of symbolic modeling for 22 commonly used String APIs, including unsupported symbolic heap representation for string arrays returned by \texttt{split()} \\
Type Conversion &
parseDouble() &
3 (2.1\%) &
No SMT-level support for string-to-double conversion predicates \\
Generic Types &
HashSet<?>.add(), HashMap<?,?>.put() &
32 (22.4\%) &
Hard to model method calls for generic objects, e.g. \texttt{Object.equals} and \texttt{Object.compareTo}  \\
Other Libcalls &
ScriptEngine.eval(), BigDecimal.setScale() &
3 (2.1\%) &
Falls back to concrete execution due to missing symbolic model \\
\bottomrule
\end{tabular}
\vspace{-1ex}
\end{table*}

%% file: Code/code-example-split.tex
\definecolor{codegray}{gray}{0.95}
\definecolor{highlight}{RGB}{255,255,160}
\definecolor{commentgray}{gray}{0.4}

\lstset{
    language=Java,
    basicstyle=\ttfamily\footnotesize,
    keywordstyle=\color{blue},
    stringstyle=\color{orange},
    commentstyle=\color{commentgray}\itshape,
    backgroundcolor=\color{codegray},
    frame=single,
    showstringspaces=false,
    escapeinside=||,
    columns=fullflexible,
    keepspaces=true,
    numbers=left,
    numberstyle=\tiny\color{gray},
    numbersep=8pt,
    breaklines=true,      
    breakatwhitespace=true
}

\begin{figure}
\begin{lstlisting}
public static boolean check_if_last_char_is_a_letter(String txt) {
    String[] parts = |\btHL{txt.split(" ")}|;
    String last = " ";
    if(parts.length != 0)
        last = parts[parts.length - 1];
    if(txt.length() != 0 && 
     txt.charAt(txt.length()-1) == ' ') last = " ";
    if(txt.length() == 0) last = " ";
    int last_char_pos =
      |\btHL{Character.toLowerCase(last.charAt(0))}| - 'a';
    return (last.length() == 1) &&
        (0 <= last_char_pos && last_char_pos <= 25);
}
\end{lstlisting}
\caption{SPF lacks sufficient symbolic modeling for the two highlighted API calls: \texttt{split} and \texttt{toLowerCase}. The converted SMT path constraints are thus imprecise. In contrast, PALM utilizes LLM's capability to interpret the semantics of commonly used APIs.}
\Description{A Java code listing for a method \texttt{check\_if\_last\_char\_is\_a\_letter(String txt)} that splits the input string on spaces, selects the last token, normalizes it, and checks whether the first character is a letter from \texttt{a} to \texttt{z}. Two API calls are highlighted: \texttt{txt.split(" ")} and \texttt{Character.toLowerCase(last.charAt(0))}.}
\label{fig:code-example-split}
\vspace{-1ex}
\end{figure}

%% file: Text/6.evaluations-userstudy.tex
\vspace{-2ex}
\subsection{User study}

To understand how PALM assists users in generating path-aware tests, we conducted a controlled user study comparing PALM with a baseline LLM-based approach. We evaluate whether PALM helps users better understand path coverage, identify redundant or missing tests, and produce concrete tests aligned with specific execution paths. We report task accuracy, completion time, and self-reported confidence for each tool. Our user study follows a counterbalanced, within-subject design. Each participant uses both tools in different orders, with tool usage and task assignment randomized and counterbalanced to mitigate learning and order effects.


\paragraph{Participants} We recruited 12 participants: 6 CS PhD students, 5 CS master’s students, and 1 CS undergraduate student. All participants were familiar with Java and had prior experience using LLMs.

\input{Table/userstudy-task}
\paragraph{Task} To evaluate how each tool supports path-aware test generation, participants were asked to generate tests for two subject programs using either PALM or a pure LLM chatbox. As a part of test generation task, they completed three targeted comprehension questions designed to assess their understanding of (1) the total number of feasible paths, (2) the equivalence between tests in terms of path exercising behavior, and (3) the ability to match a test to a natural-language path description. Table \ref{tab:userstudy-programs} and Table \ref{tab:userstudy-questions} summarize the subject programs and associated questions.

The two subject programs, \textit{CuteArray} and \textit{Triangle}, exhibit key elements of path complexity. \textit{CuteArray} contains a loop with inner loop conditions, where earlier iterations modify the array and affect later branching decisions, introducing data dependencies. \textit{Triangle} contains nested branches that sort the inputs before performing a triangle test, requiring different input orderings to fully exercise all program paths.

\paragraph{Experimental Design} 
We adopted a 2$\times$2 crossover design \cite{deljouyi2024leveraging}, where each participant used both tools (PALM and baseline) on two different programs in varied orders, evenly distributed to mitigate order bias and learning effects.

\paragraph{Experimental Procedure} Before the experiment, we provided participants with a brief overview of unit test generation and the concept of path coverage to ensure a shared baseline understanding. We then distributed a tutorial for PALM, introducing its features and explaining the user interface in detail. To familiarize participants with all key functionalities of PALM, we asked them to complete a warm-up task that involved selecting a path on the symbolic tree, observing the generated test, and composing their own test to verify its path exercisability on a simple tutorial program with four feasible paths, as shown in Fig.~\ref{fig:PALM-major-interface}.


\input{Table/userstudy-postQ}
\paragraph{Post-study Questionnaire} We used a post-task questionnaire to obtain feedback. Table \ref{tab:userstudy-postQ} lists the three questions to assess participants' perceived confidence in completing test generation tasks using PALM or LLMs. All three questions adopt a 5-point Likert scale, where 1 indicates "Not confident" and 5 indicates "Very confident."

\input{Table/userstudy-result}
\paragraph{Result}
\textbf{Accuracy.}
We define \textit{accuracy} as the percentage of correct responses to three task questions, each with known ground truth. 
Q1 asks for the total number of feasible paths under a loop bound of 2. 
Q2 asks which candidate test exercises the same path as a given reference test. 
Q3 asks which test exercises a path described in natural language.

As shown in Table~\ref{tab:userstudy-result}, participants using PALM achieved higher accuracy across all three questions: 0.92 on Q1 and 1.0 on both Q2 and Q3. The only incorrect response occurred on Q1, which assessed path enumeration, where one participant answered 6 instead of the correct 7. This participant was observed to correctly navigate the symbolic tree, which visualizes all feasible paths and allows users to inspect which ones are covered or missing, but still miscounted the total number. 

\input{Table/userstudy-resultQ-tex}
In contrast, participants using the LLM chatbox struggled to enumerate feasible paths and match tests to paths. On Q1, four participants underestimated the number of feasible paths in \textit{CuteArray} (4 vs.\ 7), often missing cases where the loop executes 0 or 1 iteration. On Q3, one participant selected an incorrect candidate after directly prompting the description and accepting the first response, leading to a mismatch at an inner condition due to overlooked data dependency. In comparison, PALM users could verify test-path alignment via the visualized execution trace.

\textbf{Completion time.} Participants using PALM and participants using the LLM chatbox completed tasks in a similar time range (5.43 vs. 5.17 minutes on average).

\textbf{Post-survey confidence.} Figure~\ref{tab:userstudy-resultQ} shows self-reported confidence across three aspects when using the two tools: full path enumeration capability (P1), avoidance of redundant tests (P2), and alignment between tests and target paths (P3).

Participants using PALM gave consistently high confidence ratings, with an average of 4.92 on P1 (close to very confident), 4.33 on P2, and 4.58 on P3 (both between fairly and very confident). In contrast, participants using LLMs reported noticeably lower confidence: 3.00 on P1 (moderately confident), 2.75 on P2 (between slightly and moderately confident), and 3.17 on P3 (close to moderately confident).

We observed that PALM helped participants better understand which paths were covered or missing (P1), enabling them to augment their test suites to cover missing cases. It also improved their understanding of how each test input exercises a different execution path (P2, P3), allowing them to identify which paths are covered. This was particularly useful in reducing redundant testing effort.

%% file: Table/userstudy-task.tex
\newcolumntype{P}[1]{>{\raggedright\arraybackslash}p{#1}}  

\begin{table}[b]
\centering
\caption{Programs used in user study.}
\Description{A table listing the programs used in the user study, with columns for subject name, lines of code (LOC), number of paths, and a brief description. The table includes two subjects: \texttt{CuteArray} (12 LOC, 7 paths), which counts divisible-by-7 triplets in a loop while modifying array elements, and \texttt{Triangle} (19 LOC, 12 paths), which sorts three side lengths and checks whether they form a valid triangle.}
\vspace{-2ex}
\label{tab:userstudy-programs}

\begin{tabular}{P{1.3cm} >{\centering\arraybackslash}p{0.6cm} >{\centering\arraybackslash}p{0.6cm} P{4.8cm}}
\toprule
\rowcolor{gray!30}
\textbf{Subject} & \textbf{LOC} & \textbf{Paths} & \textbf{Description} \\
\midrule
CuteArray   & 12 & 7  & Counts divisible-by-7 triplets in a loop, while modifying array elements. \\
Triangle    & 19 & 12 & Sorts three side lengths and checks if they form a valid triangle. \\
\bottomrule
\end{tabular}
\end{table}

\begin{table}[b]
\centering
\caption{Questions used in user study.}
\Description{A table listing the questions used in the user study, with columns for question ID, question title, and a brief description. The table includes three questions: Q1 on path enumeration within a fixed loop bound, Q2 on identifying a test that exercises the same execution path as a reference test, and Q3 on selecting a test that covers a specific path described in natural language.}
\vspace{-2ex}
\label{tab:userstudy-questions}

\begin{tabular}{>{\bfseries}p{0.55cm} P{1.7cm} P{5.4cm}}
\toprule
\rowcolor{gray!30}
\textbf{ID} & \textbf{Question} & \textbf{Question Description} \\
\midrule
Q1 & Path Enumeration & Identify the number of distinct feasible paths within a specified loop bound of 2. \\
Q2 & Same Path Test & From a list of candidate tests, identify the one that exercises the same path as a given reference test. \\
Q3 & Targeted Test & From a list of candidate tests, select a test that would cover a particular path described in natural language. \\
\bottomrule
\end{tabular}
\end{table}

%% file: Table/userstudy-postQ.tex
\newcolumntype{P}[1]{>{\raggedright\arraybackslash}p{#1}}  

\begin{table}[tb]
\centering
\caption{Post-study questionnaire on tool confidence}
\Description{A table showing the post-study questionnaire items measuring tool confidence, with columns for item ID, question title, and description. The items ask participants how confident they are that the tool achieves comprehensive path coverage (P1), avoids generating redundant tests for the same path (P2), and generates a test that exercises a specified target path (P3).}
\label{tab:userstudy-postQ}
\renewcommand{\arraystretch}{1.3}

\begin{tabular}{>{\bfseries}p{0.4cm} P{1.8cm} P{5.3cm}}
\toprule
\rowcolor{gray!30}
\textbf{ID} & \textbf{Question} & \textbf{Post-Study Question Description} \\
\midrule
P1 & Path Coverage Confidence & How confident are you in the studied tool's generated tests in terms of achieving comprehensive path coverage? \\
P2 & Redundancy Avoidance Confidence & How confident are you in the tool's capability to avoid generating redundant tests that cover the same path? \\
P3 & Targeted Test Generation Confidence & How confident are you in the tool's capability to generate a test that indeed exercises a given specific path? \\
\bottomrule
\end{tabular}
\end{table}

%% file: Table/userstudy-result.tex
\begin{table}[tb]
\centering
\caption{Comparison of participant task accuracy and task completion time using different tools}
\Description{A table comparing user-study results for two tools, PALM and a direct LLM baseline. The table reports participant task accuracy for three questions (Q1, Q2, and Q3) and the average task completion time in minutes. PALM shows accuracies of 0.92 for Q1 and 1.0 for Q2 and Q3 with an average time of 5.43 minutes, while the LLM baseline shows accuracies of 0.58 for Q1, 1.0 for Q2, and 0.92 for Q3 with an average time of 5.17 minutes.}
\label{tab:userstudy-result}
\renewcommand{\arraystretch}{1.2}

\begin{tabular}{>{\bfseries}p{1.4cm} | *{3}{>{\centering\arraybackslash}p{0.8cm}} | >{\centering\arraybackslash}p{2.8cm}}
\toprule
\rowcolor{gray!20}
\textbf{Tool} & \multicolumn{3}{c|}{\textbf{Accuracy}} & \textbf{Task Time (min)} \\
\rowcolor{gray!20}
              & \textbf{Q1} & \textbf{Q2} & \textbf{Q3} & \textbf{Avg} \\
\midrule
PALM          & 0.92        & 1           & 1           & 5.43 \\
LLM           & 0.58        & 1           & 0.92        & 5.17 \\
\bottomrule
\end{tabular}
\end{table}

%% file: Table/userstudy-resultQ-tex.tex
\begin{figure}[t]
\centering
\includegraphics[width=\linewidth]{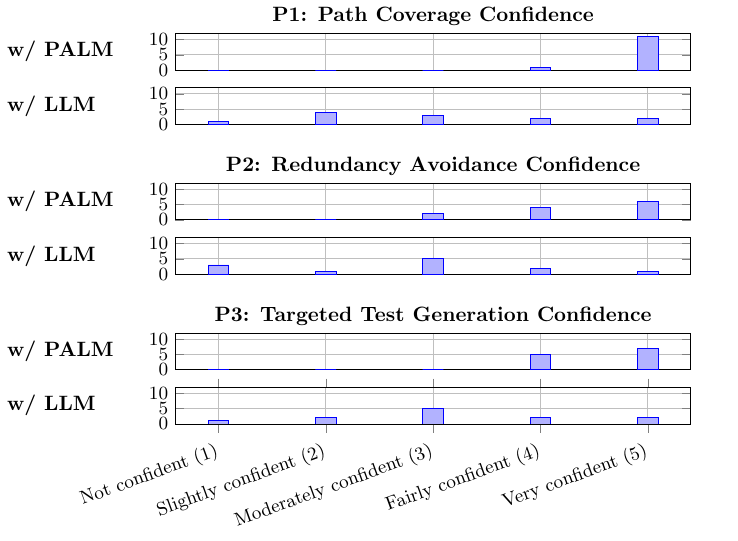}
\vspace{-3ex}
\caption{Post-study questionnaire results: distribution of confidence scores (x-axis) and corresponding participant counts (y-axis) for using PALM or LLM.}
\Description{A stacked set of bar charts showing post-study confidence ratings on a 1-5 Likert scale for three questions (P1 path coverage confidence, P2 redundancy avoidance confidence, and P3 targeted test generation confidence). For each question, two horizontal panels compare responses with PALM versus with a direct LLM baseline, with the y-axis showing participant counts (0-12). Most PALM responses concentrate at high confidence levels (4-5), while LLM responses are more spread across levels 1-5.}
\label{tab:userstudy-resultQ}
\end{figure}

%% file: Text/7.limitations.tex
\section{Threats to Validity}
\paragraph{Internal validity.}
Our comparison with direct LLM prompting is not budget-equivalent. PALM enumerates paths and performs path-specific generation and validation, which typically requires longer prompts and higher LLM usage than one-shot prompting. Thus, PALM's coverage gains may partly reflect more generation and validation opportunities. We do not include an ablation that equalizes LLM budgets, so results reflect PALM under its intended workflow rather than a cost-controlled comparison.

For Symbolic PathFinder (SPF), we align key search settings such as the same loop-unrolling bounds. However, SPF outputs satisfying assignments rather than executable unit tests and does not synthesize JUnit-style tests from these assignments, especially for arrays or heap-allocated objects. Therefore, we do not report SPF path coverage on HumanEval-Java; instead, we report how many programs yield at least one SMT-formatted path constraint as a coarse applicability indicator.

\paragraph{External validity.}
We evaluate PALM on HumanEval-Java, whose programs are relatively small and primarily use Java standard-library APIs. This setting may not fully represent real-world projects with larger codebases, richer third-party dependencies, and more complex behaviors arising from I/O, concurrency, reflection, and framework-driven callbacks. Moreover, our implementation targets Java, and the results may not generalize to other languages. Finally, HumanEval benchmarks may be subject to data contamination in LLM training corpora, which could inflate the absolute performance of LLM-based approaches.

\paragraph{Construct validity.}
We primarily use branch-based path coverage as the quantitative metric. This metric does not capture data-flow behaviors that do not manifest as distinct branch outcomes and may under-represent the strengths of traditional symbolic execution in generating boundary values within the same branch structure. Our user study complements coverage by evaluating users' ability to reason about path-level behaviors using PALM's interface. However, the study tasks focus on matching tests to paths and identifying redundant or missing tests, which may differ from end-to-end usage of test-generation tools in real projects. We also do not comprehensively measure other test-quality dimensions such as readability, naturalness, and bug-finding capability.

\paragraph{Conclusion validity.}
Our user study includes 12 participants, which limits statistical power and the strength of quantitative claims. Since LLM-based generation is stochastic, single-run results can be sensitive to sampling variance; we reduce this risk by repeating runs and summarizing aggregate results, but residual variance may still exist.


%% file: Text/8.relatedwork.tex
\section{Related Work}

Symbolic execution is a widely used program analysis technique for systematically reasoning about program behaviors and generating concrete inputs that witness property violations~\cite{king1975new,boyer1975select,king1976symbolic,howden1977symbolic}.
A key challenge is modeling system and environment interactions under the constraints of solver-based reasoning~\cite{baldoni2018survey}.
DART~\cite{godefroid2005dart} and CUTE~\cite{sen2005cute} provide concolic execution to execute external functions concretely, which can under-approximate path constraints and limit exploration when important conditions reside inside those calls.
To mitigate this, several systems introduce symbolic abstractions for common external APIs, e.g., KLEE's symbolic file system~\cite{cadar2008klee}, AEG's models for file systems, network sockets, and environment variables~\cite{avgerinos2014automatic}, and Cloud9's POSIX abstractions~\cite{bucur2011parallel}.

In comparison, PALM avoids translating path constraints into SMT formulas and therefore does not require SMT-level symbolic abstractions for external functions. However, the current implementation enumerates paths primarily by explicit branch outcomes. In contrast, mature symbolic execution engines \cite{puasuareanu2010symbolic} can additionally explore paths induced by runtime exceptions and implicit checks (e.g., nullness, bounds, divide-by-zero), as well as thread-scheduling nondeterminism in concurrent programs.

Large Language Models (LLMs) have recently gained popularity for unit test generation. Prior work mainly improves generation quality by enriching prompts or adding analysis-driven feedback. For example, ChatUniTest~\cite{xie2023chatunitest} augments prompts with focal-method context and dependencies, while Vikram et al.~\cite{vikram2023can} leverage API documentation to synthesize property-based tests. Other studies examine how prompt design affects coverage and bug finding~\cite{guilherme2023initial,li2023prompting}. More analysis-guided approaches include HITS~\cite{wang2024hits}, which uses slicing to incrementally generate tests, and TELPA~\cite{yang2024enhancing}, which refines tests using program analysis and feedback from ineffective attempts. IntUT~\cite{nan2025test} introduces explicit test intentions to guide generation, but it does not aim to exercise specific control-flow paths.

A growing line of work combines LLMs with symbolic execution or constraint reasoning. LLMSym~\cite{wang2024python} uses LLMs to assist in translating code or constraints into SMT for test generation, while Cottontail~\cite{tu2025large} uses LLMs to help satisfy SMT-formatted constraints and structured-input validity. DeFiAligner~\cite{deng2025llm} applies symbolic execution on smart-contract bytecode and uses LLMs to check consistency between symbolic summaries and documentation. SymPrompt~\cite{ryan2024code} embeds branch conditions into prompts to steer LLM reasoning, and HyLLFuzz~\cite{meng2024large} invokes LLMs to propose new inputs when fuzzing stalls. AutoEXE~\cite{li2025large} replaces SMT solving with LLM-based reasoning over code-level path constraints, but it provides limited transparency into which paths are actually exercised, especially in the presence of loops or recursion.

In comparison, PALM enumerates paths and constructs executable assertion-instrumented variants, enabling validated, iterative test generation with a visual interface to inspect path-level behavior and identify redundant or missing tests.

%% file: Text/9.conclusion.tex
\section{Conclusion}
We introduced PALM, a test generation system that combines symbolic path enumeration with LLM-guided test generation. By constructing executable path-specific variants with embedded assertions, PALM provides hints to LLM on which input to generate to target each path, bypasses the need for symbolic modeling of all invoked API functions, and enables precise test-path alignment. Its interactive front-end provides visibility into path coverage and test behavior, helping users detect missing or redundant tests. User study participants reported higher confidence in reasoning about comprehensiveness of generated tests, disambiguating missing or redundant tests, and comprehending which test inputs exercise a specific path. 

\section*{Data-availability Statement}
Our implementation and evaluation artifacts are available at \url{https://github.com/UCLA-SEAL/PALM}.

\begin{acks}
This work is supported by the National Science Foundation under grant numbers 2426162, 2106838, 2106404 and 2106420. It is also supported in part by funding from Amazon and Samsung. We want to thank the anonymous reviewers for their constructive feedback that helped improve the work.
\end{acks}